\documentclass[aps,amsmath,prd,showpacs,nofootinbib]{revtex4}
\usepackage{graphicx}

\def\l{\left}
\def\r{\right}
\def\be{\begin{equation}}
\def\ee{\end{equation}}
\def\bea{\begin{eqnarray}}
\def\eea{\end{eqnarray}}

\begin{document} 

\title{
Searching for Gravitational Waves from the Inspiral of Precessing Binary
Systems: Astrophysical Expectations and Detection 
Efficiency of ``Spiky'' Templates.}

\author{Philippe Grandcl\'ement}
\email[]{grandcle@phys.univ-tours.fr}
\affiliation{Laboratoire de Math\'ematiques et de Physique Th\'eorique, 
CNRS-UMR 6083, Universit\'e de Tours, Parc de Grandmont, 37200 Tours, FRANCE}
\affiliation{Northwestern University, Dept.\ of Physics \& Astronomy, 2145 
Sheridan Road, Evanston 60208, USA}

\author{Mia Ihm}
\email[]{mia@northwestern.edu}
\affiliation{Northwestern University, Dept.\ of Physics \& Astronomy, 2145 
Sheridan Road, Evanston 60208, USA}

\author{Vassiliki Kalogera}
\email[]{vicky@northwestern.edu}
\affiliation{Northwestern University, Dept.\ of Physics \& Astronomy, 2145 
Sheridan Road, Evanston 60208, USA}

\author{Krzysztof Belczynski}
\email[]{belczynski@northwestern.edu}
\affiliation{Northwestern University, Dept.\ of Physics \& Astronomy, 2145 
Sheridan Road, Evanston 60208, USA}
\date{December 16, 2003}

 \begin{abstract}
 
Relativistic spin-orbit and spin-spin couplings has been shown to modify the gravitational waveforms expected from inspiraling binaries with a black hole and a neutron star. As a result inspiral signals may be missed due to significant losses in signal-to-noise ratio, if precession effects are ignored in gravitational-wave searches. We examine the sensitivity of the anticipated loss of signal-to-noise ratio on two factors: the accuracy of the precessing waveforms adopted as the true signals and the expected distributions of spin-orbit tilt angles, given the current understanding of their physical origin. We find that the results obtained using signals generated by approximate techniques are in good agreement with the ones obtained by integrating the 2PN equations. This shows that a complete account of all high-order post-Newtonian effects is usually not necessary for the determination of detection efficiencies. Based on our current astrophysical expectations, large tilt angles are not favored and as a result the decrease in detection rate varies rather slowly with respect to the black hole spin magnitude and is within 20--30\% of the maximum possible values. 
 \end{abstract}

\maketitle

\section{Introduction}\label{s:intro}

Relativistic spin-orbit and spin-spin couplings can cause inspiraling binary 
compact systems containing neutron stars (NS) or black holes (BH), to precess, provided that at least one of the two spins is of significant magnitude and  misaligned with respect to the orbital angular momentum. This precession leads to a periodic change of the orbital plane orientation and therefore modify the inspiral gravitational wave (GW) signal received by ground-based detectors. Currently binary inspiral searches with interferometric 
detectors, soon expected to be at optimal sensitivity, such as Laser Interferometric 
Gravitational Wave Observatory (LIGO)~\cite{Abram92}, VIRGO~\cite{Caron97}, and GEO600~\cite{Danzm95}, rely on sophisticated signal processing techniques to extract the signal from the intrinsic noise of the instruments. For binary compact objects, the best available linear filter is the well-known {\em matched-filtering} technique~\cite{Helst68,OwenS99}. For the maximum possible signal-to-noise ratio (SNR), this method requires the best possible {\em a priori} knowledge of GW signal waveforms to be used as templates. Schematically, the basic idea is to search for correlations
between the output of the detector and the chosen family of templates. The maximum 
SNR is achieved when the signal coincides with one
of the templates. If this is not the case, the loss in SNR can be measured 
by the {\em fitting factor} (FF), with relates the actual SNR to the maximum one
by~:
 \be\label{e:def_FF}
 \l(\frac{S}{N}\r) = \mathrm{FF} \times \l(\frac{S}{N}\r)_{\rm max}, 
 \ee
 as introduced in~\cite{ApostCST94} (see also~\cite{GrandKV03,GrandK03}, for more details on FF). 

This paper represents the continuation of our group's work,  
previously presented in~\cite{GrandKV03,GrandK03} (hereafter Papers I and II, respectively). This series is devoted to the study of precessing compact
binaries and, more precisely to the effects of precession  on the 
detectability of binary inspiral signals.  If precession is not taken into account in the templates, it would cause a loss in SNR measured by $\langle {\rm FF}\rangle$ (averaged over source orientation) and consequently a decrease in detection rate by a factor 
$\langle\mathrm{FF}\rangle^3$, assuming a homogeneous space distribution of the sources\footnote{However, note that a recent study of the actual galaxy distribution in the nearby Universe shows that the slope-dependence of the number of sources within a certain distance is closer to 2.5 instead of 3~\cite{Nutzman}.}. It has been shown in Papers I and II as well as in~\cite{Apost96,BuonaCV03, PanBCV03} that the detection rate could be greatly reduced (by up to an order of magnitude). 
It has also been known for some time now that this effect is more important 
in binaries with high mass ratios~\cite{Apost96}. Therefore in this paper, we  
restrict ourselves to systems with a typical $1.4 M_\odot$ NS and 
and a stellar-mass BH of $10 M_\odot$. Let us note however, that according to 
\cite{BulikGB03}, 
the most probable system to be detected by the first generation of detectors is not of that type, but rather consists of two black holes with mass ratio close to unity. 
Our current understanding of 
the physics of the interior of NS suggests that, in a binary system, the NS is not expected to have a spin of significant magnitude~\cite{TayloW82,Wolsz91,BildsC92}. So we will assume that the most massive object, i.e. the BH, will carry all the spin. For this case where only one  object is spinning, Apostolatos {\em et al.} (1994-1995)
have derived an 
analytical approximation of 
the emitted waveform, valid up to 1.5 post-Newtonian (PN) order : the 
{\em simple precession} regime
\cite{ApostCST94, Apost95}. For simplicity and computational efficiency, we will use this approximation,
instead of integrating the full set of 2PN equations describing the dynamics of the system \cite{Kidde95}. We will see that the results are very close to the ones obtained by Buonanno {\em et al.} using more detailed signals \cite{BuonaCV03}. As in Papers I and II 
we will use the initial LIGO noise curve throughout the paper (values taken 
from table IV of \cite{DamouIS01}).

In a realistic data analysis pipeline, it is difficult to implement a search 
that uses the ``full'' precession waveforms as templates. Indeed such waveforms
depend on a large number of parameters, typically ten (e.g., magnitude and 
orientation of the spins, orientation and position of the binary, etc), so that 
the computational burden to compute the cross-correlations with the output of
the detector for all the possible values of the parameters is beyond the 
reach of current resources. To solve this problem, one can hope to find
a family of ``mimic'' templates which depend on formally few parameters, but 
still leads to high $\mathrm {FF}$ values. Following and extending an initial proposal by Apostolatos \cite{Apost96}, we presented, in Paper II, a strategy based on a hierarchical search, where successive corrections are applied to the phase of the templates. The final correction is the sum of an oscillatory term of 
the form proposed by \cite{Apost96} and several spike-like features, hence the
name ``spiky'' templates.

Apart from the mass ratio, the effect of the precession modulation and hence 
the $\mathrm {FF}$ also depends on the 
spin magnitude of the most massive object (here the BH of $10 M_\odot$) and 
the tilt angle of this spin with respect to the orbital angular momentum. 
Until now none of the studies focused on the detectability of precessing binary inspiral have considered the astrophysical origin of the compact object spin magnitudes and orientations and their expected probability distributions. Clearly these will affect any {\em realistic} probability statement about the detectability of precession-modulated signals must take these into account. The physical origin of the tilts has been understood uniquely in the context of asymmetric compact object formation and the existence of supernova kicks~\cite{KaspiBMSB96,WexKK00,PorteY98}. Our current astrophysical understanding is well developed and allows theoretical predictions for the distribution of tilt angles among different populations of binary compact objects. On the other hand, the evolution of compact object magnetic field strengths is not as well understood, and therefore reliable probability distributions of spin magnitudes are difficult to calculate and the results are strongly model dependent. 

The goal of this paper is twofold. In the first part of the paper
(Sec.~\ref{s:spiky}), we examine more realistic precession waveforms (that include Thomas precession)
 and we show 
(Sec. \ref{ss:FF_curves}) that the results are in 
excellent agreement with those obtained through numerical integration of equations that describe the orbital dynamics~\cite{BuonaCV03}. In section~\ref{ss:help} we point out our intriguing result that precession can, in some cases, help detection and increase the value of the 
signal-to-noise ratio.
In the second part of the paper (Sec.~\ref{s:tilts}) we introduce, for the first time, the coupling between astrophysical predictions for the probability distributions of tilt angles with FF calculations and we obtain astrophysically relevant results for the detection rate degradation as a function of the compact object spin magnitude.  Conclusions are presented in 
section~\ref{s:conclu}.

\section{Behavior of the ``Spiky'' templates}\label{s:spiky}

\subsection{Dependence of FF on adopted precession inspiral signal}\label{ss:FF_curves}

In Paper II, we proposed a technique to detect gravitational waves from
precessing binaries, based on a three step hierarchical search. The  first
step was to just use the standard chirp templates, possibly including  2.5 PN
corrections to the phase \cite{SathyD91,DhuraS94,BalasD94}. The second
step was originally proposed by Apostolatos \cite{Apost96} and consists  of
the addition of a three parameter ${\mathcal C, B,} \delta $  oscillatory term in the phase ${\mathcal C}
\cos\l({\mathcal B} f^{-2/3} + \delta\r)$ (see Eq.
(12) of \cite{Apost96}). Although this helped, we proposed to improve  the
efficiency of the procedure by searching for spike-like features (that  we found appear in the residual phase after the two steps) in the
phase as a third correction. Each spike depends on only two (rather well constrained) parameters  (the
position $f_0$ and the width in frequency space, given by $\sigma$)
plus a boolean parameter (the sign $\varepsilon = \pm 1$).
More precisely the equation of a given spike is~:
 \bea
 \label{e:spike}
 \mathrm{If} \quad f>f_0 \quad  \mathrm{then} \quad P\l(f_0, \sigma, 
\varepsilon\r) 
 &=& \varepsilon \pi \l[\sqrt{\l(1-\frac{1}
{\l(\sigma\left(f-f_0\r)+1\r)^2}\r)}-1\r] \\
 \nonumber
 \mathrm{If} \quad f<f_0 \quad  \mathrm{then} \quad P\l(f_0, \sigma, 
\varepsilon\r) 
 &=& \varepsilon \pi \l[-\sqrt{\l(1-\frac{1}{\l(
\sigma\l(f-f_0\r)-1\r)^2}\r)}+1\r].
 \eea
The spikes are
searched one by one until some convergence criterium is achieved. We
refer the reader to Paper II for more details.

The results presented in Paper II were quite encouraging. However  there is a number of points that still need some exploration. One of them is  an
important difference between the results presented in Paper II and the  ones
obtained in parallel and independently by Buonanno {\em et al.} \cite{BuonaCV03}.  Indeed,
even though the two proposed schemes are completely different, the  effects of
precession should be the same, when using the standard (non-precessing) chirp signal.  This was
not the case : the lower curve of Fig. 12 of Paper II shows a fitting  factor
as low as 0.5 whereas, for the same system, Fig. 14 of \cite{BuonaCV03}  exhibits
minimum values around 0.7. After some investigations, we determined  that the main
difference was coming from the assumption made in Papers I and II that
the Thomas precession term in the simple precession formalism
(given by Eq. (29) of \cite{ApostCST94}) is not important and could be  ignored. As mentioned in Paper I,
this term was believed to be mainly monotonic and well recovered by  mismatching
the parameters of the templates with respect to the physical ones of  the signal. In contrast as we will see, the addition
of the Thomas precession term (using 1.5 PN expressions for  $\dot{\omega}$)
in the computation of the signal changes significantly the FF values.

Along with this main improvement we made a few other modifications. First, as  already mentioned in Paper II, we have implemented the use of FFT  techniques to find the maximum FF with respect to the time at  coalescence $t_c$, at the first step of
our procedure. By increasing the
speed of the code, this modification enables us to explore a greater  part of
the parameter space. For example, contrary to Papers I and II, for  which we used
mainly Newtonian order for both the templates and the non-precessing  part of
the signal, in this work, we always use the 2.5 PN chirp signal, which  depends on the total mass $m_{\rm T} = m_1+m_2$ and the chirp mass ${\mathcal  M} = \l(m_1^3m_2^3/m_{\rm T}\r)^{1/5}$. We determined that using
1024 points for the FFT algorithm is sufficient to reach an accuracy of  the order of $1\%$ in identifying $t_{c}$.

We also modify some of the
computational parameters involved in setting the templates database.  More
precisely, we increased the number of templates for both $m_{\rm T}$ and
${\mathcal M}$. In the previous papers, we determined those numbers by
demanding that, in the cases {\em without} precession, the recovered FF
would be $\mbox{FF} \geq 0.97$, which is similar to what is done for  the searches with
LIGO. On the other hand, it is possible that this criterion is not be  sufficient for precise
comparison with the work by Buonanno {\em et al.} \cite{BuonaCV03}.  Indeed
in \cite{BuonaCV03}, the maximum FF is obtained by using a maximization  algorithm
and not by setting a grid of templates so that the precision is  probably better
than the one obtained by imposing $\mbox{FF} \geq 0.97$ without  precession.
In order to
make a meaningful comparison we increased the number of templates for  the
masses, requiring that the resulting precision of the FF was better  than $1\%$
, measured by convergence of the FF when increasing the number of  templates
(see Paper II for details). 
Once again, this leads to using too many templates especially compared  to real searches.
The new computational parameters for the masses are the following~: the chirp mass 
(or the total mass) parameter is chosen between $80\%$ and $115\%$ 
($0.5\%$ and $310\%$) of 
the signal masses. Those intervals are populated, in logarithmic scale, by $250$ templates
for the chirp mass and $350$ for the total mass.

The last modification is a  slight change
in the criterion used for ending the search for spikes. In Paper  II we used
a relative convergence of the FF. However, for real searches, it is probably more 
convenient to adopt a criterion with respect to the SNR, given that the FF is not a very 
useful notion for noisy signals. 
Therefore we end the  search for spikes
when the SNR is modified by less than $\Delta_{\rm SNR}=0.2$
between two consecutive spike fittings. Let us however note that this value is an important 
parameter. For real searches, in the presence of noise, preliminary results seem to indicate 
that it should be carefully chosen in order to avoid high false alarm rate. The precise effect 
of noise on the spikes is a subtle subject and is beyond the scope of this paper. Let us just 
mention that we have to deal with two competitive effects : on one hand, the smaller 
$\Delta_{\rm SNR}$, the more spikes we find and so the higher the FF. On the other hand, the 
smaller the $\Delta_{\rm SNR}$ the more spikes are found, even in the presence of only noise, 
thus leading to a high false-alarm rate. This, once again, explains that this parameter 
must be chosen very carefully, given the specific behavior of the instrument noise.

\begin{figure}
\includegraphics[height=8cm]{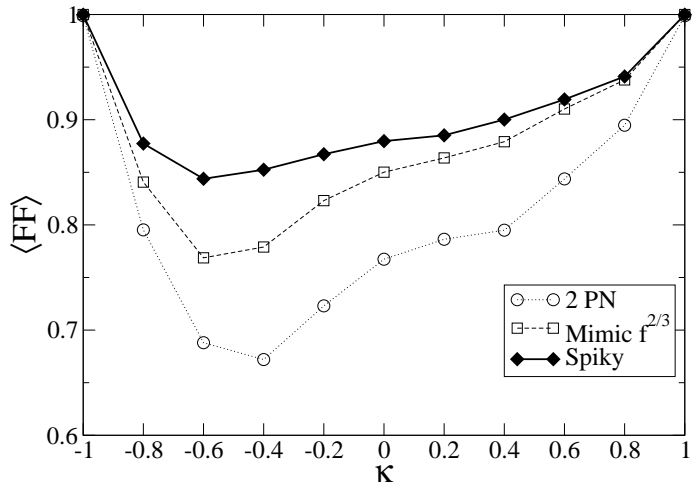}
\caption{\label{f:ff_spikes} Efficiency of the ``spiky'' templates in
recovering the signal-to-noise ratio and increasing the inspiral  detection
rate. The circles and dotted lines correspond to the values obtained
using the post-Newtonian templates. The squares and dashed lines  correspond to the 
Apostolatos' waveforms alone, and the filled diamonds and solid lines
correspond to the combination of the Apostolatos' correction and spiky templates.
Each point represents the average $\langle{\rm FF}\rangle$
for 2,000 sets of source orientations. The curves
are presented as a function of the cosine if the BH spin-orbit tilt angle : $\kappa  \equiv
\hat{\bf S} \cdot \hat{\bf L}$. We recall that the system is composed  of two bodies :
$m_1=10M_\odot$ and $m_2=1.4 M_\odot$ with only $m_1$ spinning. In this  case the spin
is maximum so that $S_1=1$. }
\end{figure}

In Figure \ref{f:ff_spikes} we show the calculated average $\langle{\rm  FF}\rangle$ values as a
function of the cosine of the misalignment angle $\kappa \equiv \hat{\bf S} \cdot \hat{\bf L}$, which is a conserved quantity, as long as simple precession is used. We consider only 
a maximally
spinning BH (the spin measured as a fraction of $m_1^2$ is $S_1=1$).
The circles and the dotted line denote the $\langle{\rm FF}\rangle$
obtained by using only the
standard templates. The most striking feature is that the curve is well
above the values found in our Fig.\ 12 of Paper II. We have verified that  two factors contribute to that : i) the use of PN templates instead of
Newtonian ones ii) the addition of the Thomas precession term, the  latter
being the more important contribution, roughly by a factor of five.
It is somewhat surprising that  adding
a term related to precession actually increases the FF, but this seems  to be
the case. The lower curve of
Fig. \ref{f:ff_spikes} is now in very good agreement with the one  labeled
``SPAs'' in Fig.\ 14 of \cite{BuonaCV03}, indicating that the simple  precession
regime is, indeed, a good approximation. Let us point out that the only
significant difference is present for $\kappa$ close to unity. This  effect may seem
surprising but is easy to understand and has actually little to do with  precession. In the
simple precession regime, when $\kappa=1$, the signal coincides exactly  with the
chirp SPA signal (i.e. the modulations ${\rm AM}$ and ${\rm PM}$ in Eq.\  (10) of
\cite{Apost95} are zero). So, for values of $\kappa$ close to unity  
(i.e., small spin tilt angles), with a signal
generated via simple precession, the FF should go to unity, as observed  in
Fig.\ \ref{f:ff_spikes}. However, when using another way of generating  the signal,
that may not be true. Indeed, in \cite{BuonaCV03}, the signal is  generated by
numerically integrating equations for $\omega$, the spin and the
orbital momentum. In the case $\kappa = 1$ the only relevant equation is
the one for $\omega$ (i.e. Eq. (1) of \cite{BuonaCV03}). The numerical  solution
coincides with the one used to derive the SPA waveform (see Eq. (7.11a)  of
\cite{WillW96} for example), only up to the given PN order. Higher  order terms
cause a small difference between the two waveforms, difference  sufficient
to explain an average FF as low as 0.8, even for $\kappa =1$. All those
considerations illustrate the very important fact that the detailed  results can
depend greatly from the exact formalism used to generate the precessing
signal. Although we will restrict ourselves
to simple precession in this paper, we plan
to conduct a more extensive study of those effects in the future.

Figure \ref{f:ff_spikes} shows that, for low values of the spin tilt  angle (i.e., $\kappa > 0.6$), $\langle{\rm FF}\rangle$ is close to 0.9  or higher. With such high values,
the spikes are not as important in recovering the precession signal  (this is why the solid and dashed curves almost coincide). However, for  larger values of
$\kappa$, the oscillatory correction begins to fail and the addition of  the
spikes leads to a significant gain, the maximum efficiency of the spikes
being around $\kappa = -0.5$. After the three steps of the procedure,  we are
able to obtain fitting factors around 0.9, for all the possible values  of
the spin tilt angle. Such efficiency is comparable to
the one shown by the alternative technique proposed by Buonanno {\em et  al.}
(see the top solid curve of Fig. (14) of \cite{BuonaCV03}).

\begin{figure}
\includegraphics[height=8cm]{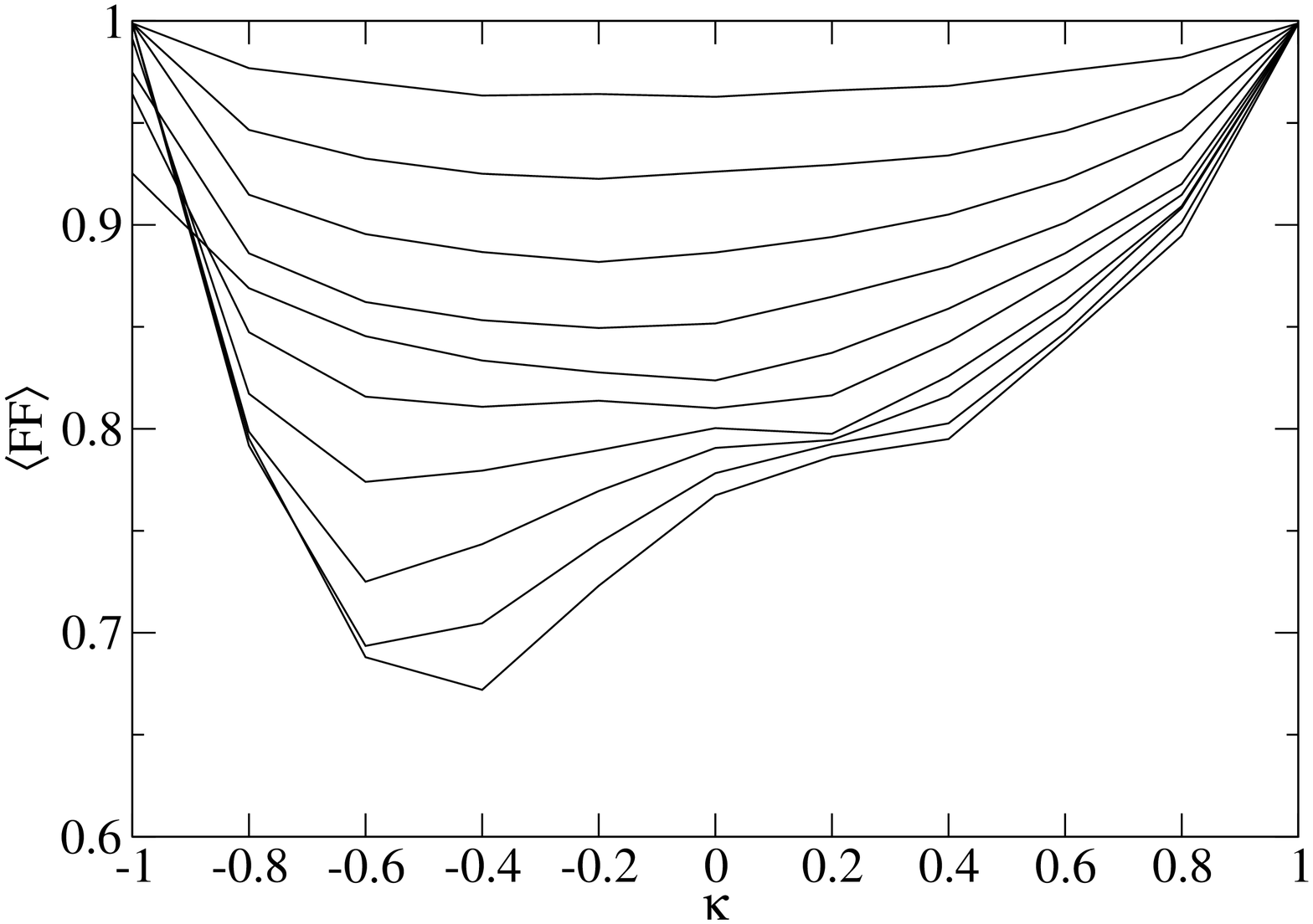}
\caption{\label{f:ff_curves}
$\langle{\rm FF}\rangle$ computed with the
non-precessing templates, as a function of $\kappa$. The spin
magnitude ranges from $0.1$ (highest curve) to $1$ (lowest curve), with a step of $0.1$. Each point is the average $\langle{\rm FF}\rangle$ for 2,000 sets of source orientations.
}
\end{figure}

In order to convolve the results  with astrophysical expectations (see Sec.\ref{ss:astro}), we have computed $\langle {\rm FF} \rangle$ values for the 
standard, non-precessing 2PN templates for a wide range of BH spin magnitudes.
 Results are shown  in Fig.\
\ref{f:ff_curves}. The spin magnitude ranges from $0.1$ (upper curve)  to $1$ (lower curve),
with an increment of $0.1$ from curve to curve.
The computational parameters are the same
than the ones used for Fig \ref{f:ff_spikes}.
Let us point out that the dependency on $S$ is
very moderate, at least for small tilt angles (i.e. high values of  $\kappa$). The curves
that do not go to $1$ when $\kappa = -1$ (see $S=0.5$ for example)  correspond to cases
where the simple approximation breaks down. This is not a worrisome  effect, given that
such break-down only occurs for very misaligned configurations  ($\kappa$ close to $-1$)
which are very unlikely to occur in reality (see Sec. \ref{ss:astro}).

\subsection{Can Precession Help Detection ?}\label{ss:help}

\begin{figure}
\includegraphics[height=8cm]{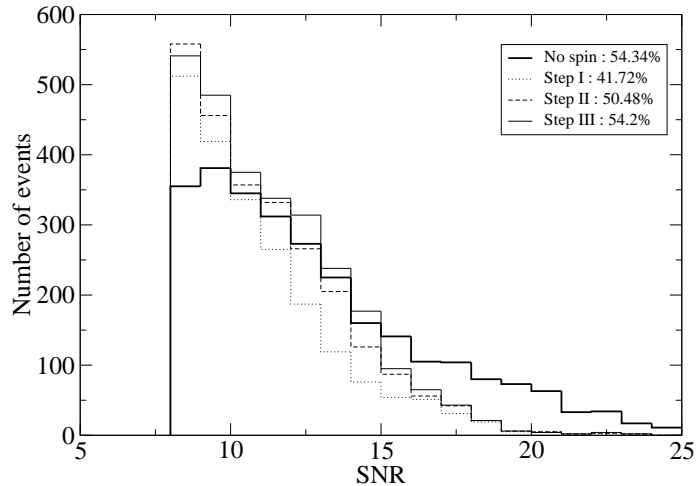}
\caption{\label{f:histo} Histogram of the signal-to-noise ratios at  which
the detection of 5,000 different orientations of a binary $m_1=10  M_\odot$,
$m_2=1.4 M_\odot$ and $S=1$ is achieved. The dotted line denotes the  results
using the non-precessing standard templates, the dashed one after  including
the oscillatory term to the phase and the solid one after the addition  of
several spikes. The thick curve labeled {\em no spin} denotes the  values of SNRs
that are achieved by setting the spin of the BH to zero, keeping all  the other
parameters fixed.
Only the detected events
(i.e. those with ${\rm SNR} \geq 8$ are shown). The
percentiles indicates the fraction of detected events with respect to  the total
number of binaries.
}
\end{figure}

The efficiency of our procedure can also be presented without using  the concept of FF, 
but rather by computing SNRs.
Fixing the distance to $20 \mbox{Mpc}$, as an example,
we simulate a set of 5,000 different binaries. Then we compute the SNR  at which detection is achieved. Figure \ref{f:histo} shows the number  of events
for each particular interval of SNRs. The three different curves show  the values
after each step of our three-step procedure. Only the events with ${\rm  SNR}
\geq 8$ are presented (as this is our assumed detection limit). We can  see that, even though the overall shape of the distribution
does not change drastically, more and more events are detected when the  successive
corrections to the phase are added. The fraction of detected events are  given
by the percentiles in the legend-box of Figure \ref{f:histo}. The  improvement coming
form Apostolatos' ansatz (Step II) and from the spikes (Step III) are both 
significant, the latter being half of the first one.
Given the results of Fig. \ref{f:ff_spikes}, we expect that the improvement 
comes mainly from
systems with $\kappa$ close to $-0.5$ (remember that, in this  section,
$\kappa$ is evenly distributed in $\l[-1,1\r]$ ; for astrophysical  considerations
on the distribution function of $\kappa$, see Sec. \ref{s:tilts}).

The thick black curve of Fig. \ref{f:histo} represents the results using the same sets of
orientations but setting $S=0$. We first point out that the shape of the  distribution is
somewhat different from the same cases with precession (dotted line of Fig. \ref{f:histo}).
On one hand, for signals without  precession there is a larger number of
signals with high SNRs compared to those with precession, and this is  quite intuitive. On
the other hand, for low to moderate SNRs ($8$ to $10$), there is  significantly
more systems with precession that without. This effect is so important  that the number of events detected with precession, after the last step ($54.2\%$) 
is very close to the one without precession ($54.34\%$), even with FF values 
smaller than 1.

The conclusion is that there are systems for which the SNR 
with precession is higher than  without.
How is that possible ? Let us recall that the actual SNR is given by Eq.
(\ref{e:def_FF}), where the $\l(S/N\r)_{\rm max}$ is given by the norm  of the
signal, i.e. by $\l(W|W\r)^{1/2}$ (the optimum choice of scalar product  is given by
Eq. (2.3) of \cite{CutleF94} ; see also  \cite{Apost96,Apost95,Helst68,Finn99} for
details). Without precession the FF is 1 so, in order to get systems  with
higher SNR when $S=1$ than when $S=0$, there must exists some systems  for which
$\l(S/N\r)_{\rm max} \l(S=1\r) > \l(S/N\r)_{\rm max} \l(S=0\r)$. In  other words,
at least for some systems, precession will {\em enhance} the strength  of the
gravitational wave signal. This can be physically understood as  follows: consider a system which, when entering the
frequency band of the detector, is
oriented in the worst possible configuration, hence giving the lowest  SNR
possible. In the case $S=1$, because of precession,
the orientation of the orbital plane with
respect to the detector is going to change and the system will no  longer be
in the worst possible orientation, hence will have higher SNR that what  would be
the case without precession. Of course there must also be systems for  which
the reverse is true, i.e. for which precession decreases the SNR.

\begin{figure}
\includegraphics[height=8cm]{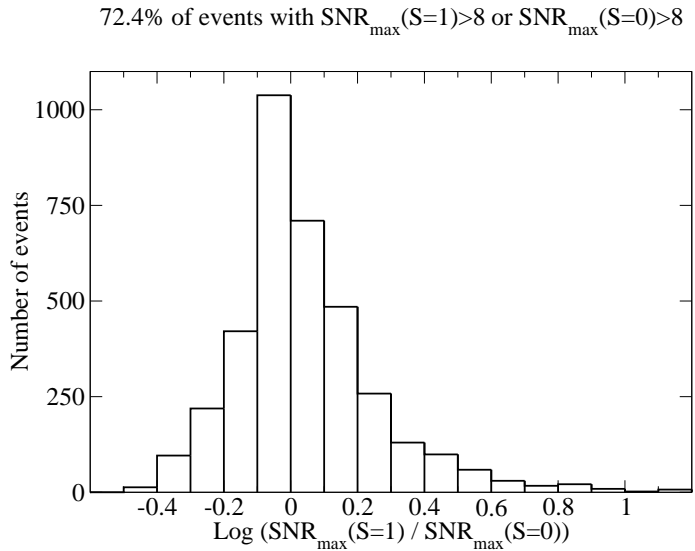}
\caption{\label{f:SN_max}
For events that are detected at least in one of the two cases $S=0$ or  $S=1$, this
histogram shows the distribution of the ratios of the two  $\l(S/N\r)_{\rm max}$, in
logarithmic scale. This illustrates the fact that, depending on the  actual
orientation of the system, precession can either increase or decrease  the strength
of the signal.
}
\end{figure}

To be more
quantitative, we compute $\l(S/N\r)_{\rm max}$ with $S=0$ and $S=1$ for  a set
of 5,000 configurations. Then we isolate the $72.4\%$ of the systems  that are
detected in one case or the other, i.e. the systems for which either
$\l(S/N\r)_{\rm max}\l(S=0\r) \geq 8$ or $\l(S/N\r)_{\rm max}\l(S=1\r)  \geq 8$.
For those systems, Fig. \ref{f:SN_max} shows the distribution of the  ratio
of the two $\l(S/N\r)_{\rm max}$,
in logarithmic scale. As expected, there is roughly
half of the configurations for which precession enhances the strength of  the signal
and half for which it decreases it. However, let us note that the distribution is 
asymmetric so that the mean increase value is higher than the mean decrease one.
This modification accounts for the  fact that
the distribution of SNR on Fig. \ref{f:histo} is different in the two  cases
$S=0$ and $S=1$ (respectively thick black curve and dotted curve).
So to conclude with  this effect, the ``real'' SNRs are modified
by precession in two ways~:
\begin{itemize}
\item the match between the templates and the signal is no longer  perfect. This
is measured by a decrease in FF.
\item the $\l(S/N\r)_{\rm max}$ are modified, which can either increase  or
decrease the strength of the signal.
\end{itemize}
In view of this understanding, let us come back to Fig. \ref{f:histo}.  The second
effect is the same no matter what templates are
used. The only difference between the various steps of our procedure is  the
fact that the FF increases. It is interesting to note that when using  only the first
two steps (non-precessing templates and Apostolatos' ansatz) the net  effect of
precession is to reduce to number of systems detected. However, when  the spikes
are included, the decrease of FF is sufficiently small to be  compensated by the
enhancement in $\l(S/N\r)_{\rm max}$, causing roughly the same number of 
 systems to be  detected than in the case $S=0$.
However, it is true that for some events the
$\l(S/N\r)_{\rm max}$, hence the actual SNR, decreases, as
it seems to be the case for the systems with very high SNRs. To  summarize, we have
illustrate the fact that, by modifying the signal, precession, if  correctly
accounted for by the templates (i.e. with sufficiently high FF),
can help recovering the same number of detection than with $S=0$.
From this point of view, it  seems that, in some cases, precession 
could  help the  detection of inspiral signals.

\section{Spin Tilts in Double Compact Objects}\label{s:tilts}
\subsection{Spin Tilt Astrophysics}\label{ss:astro}

 Inspiraling binaries have already been detected in our Galaxy and are known as relativistic binary pulsars (the Hulse-Taylor systems PSR~B1913+16 being the prototype~\cite{HulseT75}). All the known double compact objects with coalescence lifetimes shorter than the Hubble time consist of two neutron stars or a neutron star and a white dwarf (their inspiral is relevant to the planned space mission LISA~\cite{LISA_NASA,LISA_ESA}). Nevertheless current theoretical understanding of binary compact object formation (consistent with the binary pulsars) undoubtedly leads to the prediction for the existence of close binaries with a NS and a BH, and for the majority of models of binary BHs (e.g.\cite{BelczKB02} ; hereafter BKB) . Their formation is understood as the outcome of many different binary evolutionary sequences all of which involve a long chain of phases of mass and angular momentum transfer between binary components (the progenitors of the compact objects) as well as mass and angular momentum losses from the binaries~(see BKB). Of all these evolutionary links there is only one that is believed to be the origin of spin-orbit misalignments, at least for binaries in the galactic field, i.e. not in globular clusters: the asymmetric collapse of massive stars that imparts a randomly oriented recoil velocity (known as ``kick'') to the nascent compact objects, which affects the binary orbital characteristics and depending on its magnitude and direction can cause the orbital plane to tilt from its pre-collapse orientation. Observational evidence in support of NS kicks has accumulated over the past two decades from a variety of sources, most important being the large measured space velocities of 
pulsars~\cite{HeuveP97}. For a couple of binary pulsar systems (one of them being the Hulse-Taylor binary), careful timing and polarization pulsar measurements have provided evidence in support of a non-zero tilt angle between the pulsar spin axis and the orbital angular momentum axis. Such misalignments have been theoretically interpreted as the result of asymmetric kicks and have been used to provide quantitative constraints on kick magnitude and orientation~\cite{KaspiBMSB96,WexKK00}. 

This basic theoretical understanding has already been used in a preliminary study of the tilt angle distributions expected for BH-NS and BH-BH binaries and their general implications for the reduction in detection rate, if non-precessing templates are used in GW inspiral searches (see Kalogera \cite{Kalog00}, hereafter K00). As discussed in K00 in more detail, the tilt-angle calculation is possible in part because of two main evolutionary factors: 
 \begin{itemize} 
 \item{The complicated evolutionary history of binary compact objects includes multiple phases of strong tidal interactions and mass-transfer phases. The spins of the binary components are expected to be aligned with the orbital angular momentum axis just prior to the formation of the second compact object because of the exchanged angular momentum between the stars and the orbit through these tidal interactions. Therefore, even if earlier in the evolution of the binary compact object progenitors spins were misaligned, all momenta axes are expected to be aligned just prior to the second supernova explosion.} 
 \item{In a BH-NS binary the BH is expected to form from the initially more massive star in the binary. Since more massive stars evolve faster, the BH is expected to be formed first and its spin is expected to be aligned with the orbit just prior to the second explosion. The NS formed in this second explosion is expected to receive a natal kick, which can result in the post-explosion orbital plane being tilted with respect to the pre-explosion orbital plane. Given the extremely small cross section of the BH, its spin remains unaffected and aligned with the pre-explosion orbital angular momentum, and hence {\em misaligned} with respect to the orbital angular momentum axis of the newly formed double compact object\footnote{Similar considerations hold for a BH-BH binary, where the NS is typically replaced by the least massive of the two BHs. However, note that it has been recently shown \cite{BulikGB03} that relatively massive BH-BH with mass ratios close to unity have the highest detection probability and that such systems are not expected to receive significant kicks (if any).}.}
 \end{itemize}
 K00 developed a basic scheme for the quantitative modeling of the orbital dynamics during an asymmetric explosion and the resultant distributions of orbital-plane tilt angles, under the assumption of {\em circular} pre-supernova orbits. Based on the above considerations these tilt angles are the misalignment angles between the BH spin and the orbital angular momentum axis, right after the formation of the double compact object. Furthermore, it has been shown that the time-scale for any evolution in the tilt angle due to spin-spin and spin-orbit couplings is much longer than the age of the universe~\cite{Ryan95}. Consequently, the post-explosion tilts are expected to be preserved during the inspiral phase and to be relevant at the very late phases when the GW frequencies enter the detection bands of ground-based interferometers ($\gtrsim 40$\,Hz for initial LIGO, for example). The results in K00 indicated that BH-NS systems can have significant (up to $70\%$ of the systems with tilt angles in the range 
$50^\circ-100^\circ$) tilt angles, whereas heavier BH-BH systems with typically smaller kick magnitudes tend to have very small tilt angles (more than $90\%$ of the systems with a tilt angle smaller than $30^\circ$). 

Apart from the assumption of circular pre-explosion orbits, K00 adopted a number of simplifications: (i) tilt distributions were calculated for a few specific sets of pre-explosion binary parameters (component masses and orbital separations) without taking into account the probability distributions of pre-explosion binary characteristics, given the prior evolutionary history of the systems; (ii) the tilt distributions were not convolved with detailed calculations of FF and its dependence on tilt angle and spin magnitude. Here we present a more sophisticated analysis that expands the preliminary study in K00 in three major ways: (i) we use binary compact object population models that account for the full evolutionary history of binaries from their formation as primordial binaries in the Galaxy until double compact objects form; these population models provide a detailed record of the distribution function of binaries in their parameter space (4-dimensional: two masses, orbital semi-major axis, and eccentricity) just prior to the second explosion that introduces the BH spin tilt; (ii) we expand the mathematical derivation of K00 to account for the general case of eccentric binaries; (iii) we convolve our astrophysical predictions for tilt angle distributions with our results for the FF values as a function of spin tilt and magnitude. This allows us to derive astrophysically ``weighted'' FF as a function of spin magnitude and examine the systematics of our results for different population models. 

\subsection{Spin Tilt Calculations and Results}\label{ss:calc}

Although many of the details of binary evolution are not well understood, the current general understanding (based both on theoretical stellar evolution calculations and comparisons with a very wide range of different binary types) is quantitative enough to allow the statistical modeling of binary populations and the calculation of probability distributions of all physical parameters of interest. For a wide variety of scientific problems, such {\em population synthesis} calculations have been used for more than 20 years now, especially in the area of binaries with compact 
objects~\cite{DeweyC87}. Most commonly Monte Carlo methods are used to generate a large number ($10^6-10^7$) of primordial binaries with physical properties that follow certain distribution functions. The evolution of each binary is independent of the other (since in the Galactic field the interaction time is longer than the age of the Milky Way) and is modeled through long sequences of evolutionary phases, until the end of the binary's lifetime. Throughout the calculation the binary characteristics are calculated self-consistently. The overall normalization of the models is fixed by observational constraints on the supernovae and the star formation rates of the galaxy.

For our calculations we use the code {\em StarTrack} that was originally developed by Belczynski et al.\ (2002) specifically for the study of binary compact object coalescence rates and physical properties (BKB). The code, results, and comparisons with earlier work by other groups have been described extensively in BKB, so we will not repeat the astrophysical description here. We further note that in the past year the population synthesis code has been modified with special focus on the treatment of mass-transfer phases:  a new, computationally efficient scheme has been developed that, for the first time allows the calculation of mass-transfer rates and binary evolution through mass-transfer phases that have been carefully tested against results of detailed stellar evolution codes  and are in very good agreement~\cite{BelczKRT03}. Using this updated code we have studied the properties of  double compact objects using a subset of the set of models considered by BKB. We find that the results from the updated version are qualitatively identical to the early models (presented in detail in BKB) and, in terms of the tilt-angle distributions, quantitatively similar but with noticeable differences (typically the new models have a somewhat larger fraction of binaries formed with tilt angles in excess of $10^\circ$). 

In what follows we consider the results from two models already presented in BKB: the standard model 
A and the model E1 (with $\alpha_{\rm CR} \lambda = 0.1$). These have been selected because they provide the widest variations in results in terms of tilt-angle distribution functions. The details of the physical parameters in each model and the reasons for their choice are described in detail in BKB. We avoid the repetition here, since, as we will see, the astrophysical motivation and details are not important for the results and conclusions from this study. However we do describe here the calculation of the tilt angles for {\em eccentric} pre-explosion orbits, since the derivation does not appear anywhere else in the literature. 

\begin{figure}
\includegraphics[height=6cm]{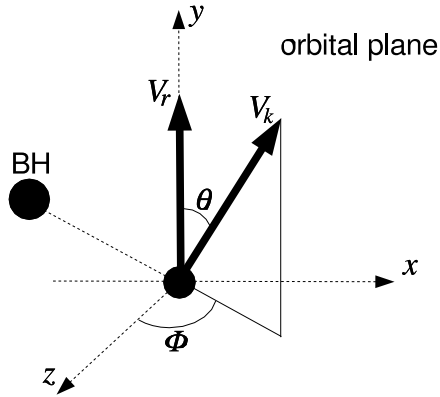}
\caption{\label{f:orient}
Orientation of the binary. The $\l(x,y\r)$ plane is the orbital one, $\vec{e}_y$ being
in the direction of $\vec{V}_r$ and $\vec{e}_z$ in the one of the orbital momentum. 
The orientation of the kick velocity $\vec{V}_k$ is given by the two angles 
$\theta$ and $\phi$. Note that, contrary to Fig. 1 of K00, the BH is no longer in the 
x-direction in the case of eccentric orbit.
}
\end{figure}

We adopt a reference frame analogous of the one used in K00. However, we need to 
slightly modify it to take into account eccentric orbits. The axes are defined as follows.
$\vec{e}_y$ lies into the direction of the initial velocity $\vec{V}_r$ of the 
progenitor, $\vec{e}_z$ in the direction of the orbital momentum and we have
$\vec{e}_x = \vec{e}_y \times \vec{e}_z$. Then, the $\l(x, y\r)$ plane is the orbital one.
 This system is the same than the one used in 
K00 except that, for eccentric orbits, the BH is no longer always in the x-direction.
Let us call $A_0$ the semi-major axis of the orbit and $r$ the separation at the 
moment of explosion. We call $V_c$ the value of $V_r$ when $r= A_0$ (the subscript 
c stands for circular). The magnitude of the kick velocity is given by $V_k$ and 
its orientation by the two angles $\theta$ and $\phi$ (cf. Fig. \ref{f:orient}). 

The velocity $\vec{V} = \vec{V}_r + \vec{V}_k$ of the NS, 
after explosion is then given by (see also \cite{Hills83}, by replacing 
$\phi \rightarrow \pi/2-\phi$)~: 
 \begin{eqnarray}
 \frac{V_x}{V_c} & = & u_k \sin \theta \sin \phi \\ 
 \frac{V_y}{V_c} & = & (V_r/V_c) + u_k \cos \theta = 
	\sqrt{2A_0/r - 1} +  u_k \cos \theta \\ 
 \frac{V_z}{V_c} & = & u_k \sin \theta \cos \phi 
 \end{eqnarray} 
where we have defined $u_k \equiv V_k/V_c$. 

 As in K00, the tilt angle $\omega$ can be calculated by 
considering that it is the angle between
two vectors: the initial velocity $\vec{V}_r$ and the projection $\vec{V}_p$ of 
$\vec{V}$ onto the plane $\phi=0$ (note a typo in K00).
This leads to~: 
 \be
\kappa \equiv \cos\omega = \frac{\sqrt{2A_0/r - 1} + u_k \cos \theta}{\sqrt{(\sqrt{2A_0/r - 1} + u_k \cos \theta)^2 + (u_k \sin \theta \cos \phi)^2}}
 \ee

In the case of a circular orbit, for which $A_0=r$, we recover Eq.\ (3) of K00.
The radius $r$ is chosen 
proportionally to the time spent at each point of the orbit before the explosion.

\begin{figure}
\includegraphics[height=6cm]{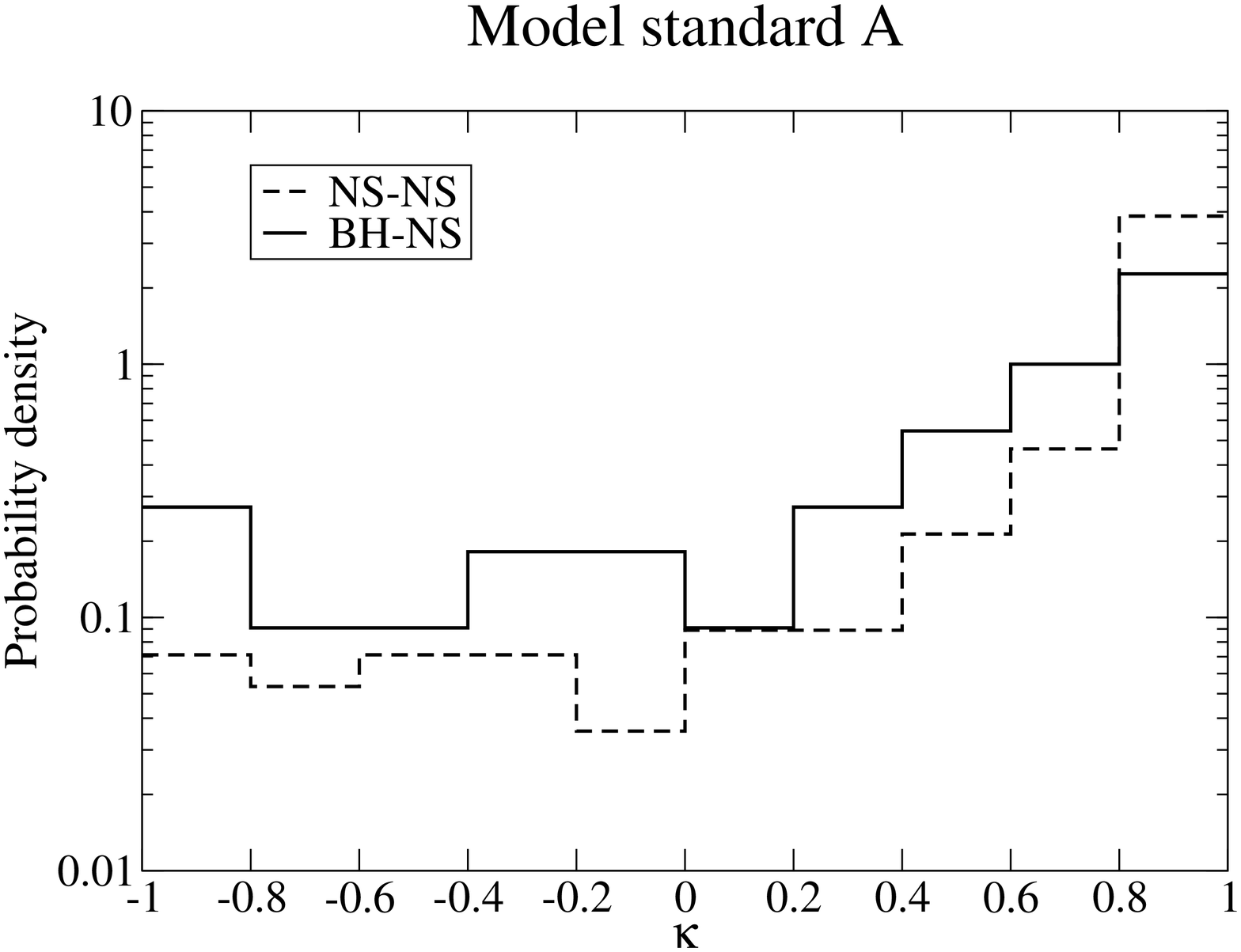}
\includegraphics[height=6cm]{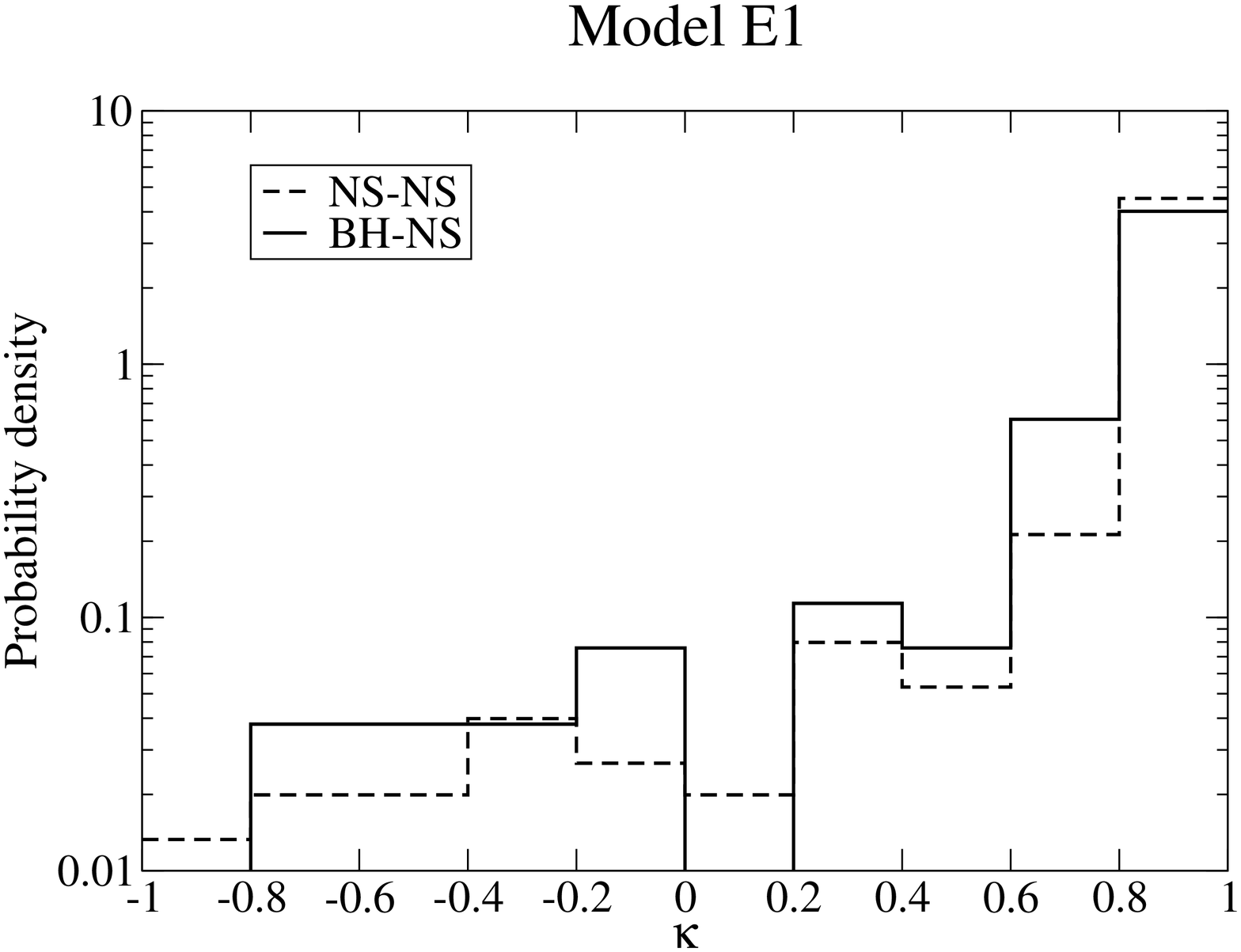}
\caption{\label{f:density} For the models A and E1, the probability distribution of the tilt-angle 
$\kappa$ is shown. On both plots, the solid curve denotes the results for BH-NS systems and the dashed 
one the NS-NS systems.}
\end{figure}

In Fig. \ref{f:density} our results for the probability distribution of $\kappa$ are shown for the two population models. Each panel shows the distribution for binaries of types NS-NS and BH-NS. We did not show the BH-BH binaries because the statistics we got was to low to be meaningful. In general it is evident that it is very difficult to tilt the orbital plane by large angles. It can be shown mathematically that the tilt angle increases as the magnitude of the kick velocity component perpendicular to the pre-explosion orbital plane ({\bf $\vec{V}_{\rm k}\cdot \vec{e}_z$}) increases too. However, as this quantity increases, statistically so does the total kick magnitude. On the other hand, the latter is has the effect of disrupting the binary and hence {\em not} forming a binary compact object. This is the physical reason for why low tilt angles are more favored: they occur in systems that have avoided disruption. 

The results from the two systems behave in a rather similar manner. The probability density 
drops rapidly of 1-2 orders of magnitude between $\kappa = 1$ and $\kappa = 0$ and remains 
relatively constant afterwards ($\kappa < 0$).
The main difference between the two models is that it is easier to get large tilt angles in the 
standard model A than for E1. This effect is very noticeable and, as we will see, will reflect on 
the $\langle{\rm FF} \rangle$ distribution. For both models, it seems that the BH-NS systems are 
slightly easier to tilt than the NS-NS ones. This is another reason, apart from the fact that 
precession is more important for system with very different masses, to consider only precession 
for BH-NS systems.

\subsection{Astrophysically Relevant Fitting Factors}\label{ss:astro_FF}

In Sec.\ \ref{ss:FF_curves} $\langle{\rm FF}\rangle$ are calculated for a representative BH-NS binary and for a grid of BH spin magnitudes $S$ and spin tilts $\kappa$. Here we use the astrophysical model predictions  for the distribution of $\kappa$ to derive {\em ``astrophysically weighted''} fitting factors $\langle{\rm FF}\rangle_{\rm astro}$ that depend only on the BH spin magnitude and are defined as: 
 \begin{equation}
 \langle{\rm FF}\rangle_{\rm astro}(S) = \int_{-1}^{1} f(\kappa) \langle{\rm FF}\rangle (\kappa,S) ~d\kappa. 
 \end{equation}

\begin{figure}
\includegraphics[height=8cm]{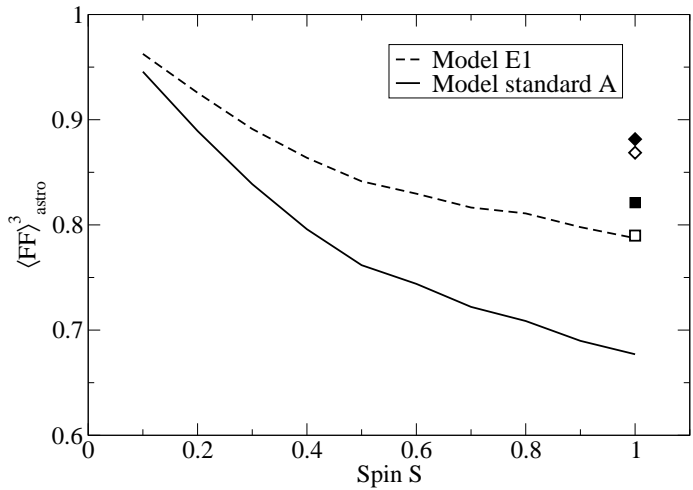}
\caption{\label{f:ff3_astro} Reduced detection rate, given a probability distribution for $\kappa$. 
The system is composed of a black hole and a non-spinning NS. $\langle {\rm FF}\rangle^3_{\rm astro}$
is presented for both models, as a function of the spin of the black hole. The
empty square (diamond) denotes the value for Apostolatos' templates, for model 
A (E1) and the filled square (diamond) the value using the ``spiky'' 
templates for model A (E1).}
\end{figure}

In Fig. \ref{f:ff3_astro} we show the derived astrophysical $\langle {\rm FF}\rangle_{\rm astro}^3$ values as a function of the BH spin magnitude. The variation across the range is moderate and $\langle{\rm FF}\rangle^3_{\rm astro}$ values are typically {\bf $\geq 0.7$}. They depend only moderately on $S$ and on the details of the population models. As expected from the probabilities shown on Fig. \ref{f:density}, the 
model E1 exhibits values above the standard one by about 20\%. The decrease in detection rate, 
if precession is ignored in the inspiral templates, is only moderate and about 20-30\%. This result is the outcome of a balance between the fact that the lowest $\langle {\rm FF}\rangle$ occur for $\kappa\sim -0.5$ (see. Fig. \ref{f:ff_curves}) 
and that there is only a low probability for BH-NS to have such large tilts 
(see Fig. \ref{f:density}). Figure \ref{f:ff3_astro} also 
shows the values obtained
when using both Apostolatos' waveforms (empty symbols) and the ``spiky'' 
templates (filled symbols), for model A (the squares) and E1 (the diamonds). 
We can see that the improvement induced by the
spikes is not very important in this case. This is not surprising, given that
the spikes are efficient in the region $\kappa \simeq -0.5$ which is a value
which is not favored at all by astrophysical models 
(see. Fig. \ref{f:density}). The inclusion of the mimic templates is also more
efficient for the model A than for E1 and is easily explained by the fact that 
the model A produces significantly more systems with large values of the 
misalignment angle.

\section{Conclusions}\label{s:conclu}

The first part of this paper is the natural continuation of previous work and is devoted 
to a more detailed study of various precessing effects. We used more detailed signals than in 
Papers I and II, always generated by the simple precession formalism, but using 2PN expressions 
for the non-precessing terms and including Thomas precession. It turns out that those two 
modifications significantly increases the values of $\langle {\rm FF} \rangle$, compared 
to what was previously found, which is somewhat surprising and counter-intuitive. On the other 
hand, the results are now in very good agreement with the ones obtained by \cite{BuonaCV03}. 

Using those more detailed signals, we then explored the efficiency of the family of ``spiky'' 
templates proposed in Paper II. The values of $\langle {\rm FF} \rangle$ being higher, 
the spikes are less efficient that previously found. However, the overall procedure still leads 
to values of $\langle {\rm FF} \rangle$ comparable to the ones found by \cite{BuonaCV03} (see 
however recent work in \cite{PanBCV03}, where even higher $\langle {\rm FF} \rangle$ are 
achieved). Using our ``spiky'' templates we then computed ``real'' SNRs, fixing the distance 
of the binary to $20 {\rm Mpc}$. In doing so, we were able to illustrate and quantify the fact that 
precession can, in some cases, help detection. This is due to the fact that, even if ${\rm FF} < 1$, 
the strength of the signal can be enhanced by precession. In future work, we plan to see 
how the efficiency of our procedure is modified by using more complicated signals, mainly 
by dropping the adiabatic approximation. We also plan to explore the behavior of the 
``spiky'' templates in the presence of noise.

In the second part of this paper, for the first time, we explore the effect of precession on gravitational-wave inspiral searches in view of the most current understanding of the physical origin of spin-orbit tilt angles and associated predictions for the probability distributions of tilt angles. Taking into account detailed models for the full evolutionary history of close binary compact objects we calculate the probability distributions of tilt angles for different types of systems and find that the strong majority of tilts are smaller than $\simeq 60^{\circ}$. This is mainly due to the requirement that binaries remain bound after compact object formation {\em and} in a tight orbit (with coalescence times shorter than the Hubble time). 

We use the predictions for tilt-angle distributions to calculated ``astrophysically-weighted'' fitting factors $\langle{\rm FF}\rangle_{\rm astro}$ as a function of black hole spin magnitude. We find that most probably Nature protects us from a large decrease of the inspiral detection rate in the case that non-precession templates are used: for a wide range of binary evolution models, the decrease factors of detection rates turn out not to depend strongly on the spin magnitude and they remain rather well restricted within $20-30$\% of the maximum possible. 

The implication of our results on the astrophysical expectations of tilt angles is that precession effects may not as important as previously thought for the searches of gravitational-wave inspirals. However, precession in principle can still be important, if larger tilt angles are more favored than currently thought. Given the fact that the current understanding of binary compact object formation is far from perfect and that the anticipated detection rates for first-generation interferometers may be low, it is probably best to still account for precession effects in the best way possible in gravitational-wave data analysis.

\begin{acknowledgments}
This work is supported by NSF Grant PHY-0121420. V.K. also acknowledges financial support 
form the David and Lucile Packard Foundation, in the form of a Fellowship in Science and 
Engineering.
\end{acknowledgments}

\end{document}